# Bus Protocols: MSC-Based Specifications and Translation into Program of Verification Tool for Formal Verification

Kamrul Hasan Talukder

Computer Science and Engineering Discipline
Khulna University, Khulna 9208, Bangladesh.
E-mail: kamrul9375@yahoo.com

*Abstract*

*Message Sequence Charts (MSCs) are an appealing visual formalism mainly used in the early stages of system design to capture the system requirements. However, if we move towards an implementation, an executable specifications related in some fashion to the MSC-based requirements must be obtained. The MSCs can be used effectively to specify the bus protocol in the way where high-level transition systems is used to capture the control flow of the system components of the protocol and MSCs to describe the non-atomic component interactions. This system of specification is amenable to formal verification. In this paper, we present the way how we can specify the bus protocols using MSCs and how these specifications can be translated into program of verification tool (we have used Symbolic Model Verifier (SMV)) for the use of formal verification. We have contributed to the following tasks in this respect. Firstly, the way to specify the protocol using MSC has been presented. Secondly, a translator that translates the specifications (described in a textual input file) into SMV programs has been constructed. Finally, we have presented the verification result of the AMBA bus protocol using the SMV program found through the translation process. The SMV program found through the translation process can be used in order to automatically verify various properties of any bus protocol specified.*

## 1. Introduction

Message Sequence Charts (MSCs) are an attractive visual formalism used in the early design stages of systems to capture system requirements. MSCs and a related mechanism called High-level Message Sequence Charts (HMSCs) have been standardized [1] for specifying telecommunication software. A version of MSCs called Sequence Diagrams are also a behavioral diagram type used in the UML standard [2]. These uses of MSCs are mainly in capturing the system requirements. However, if we move towards an implementation, an executable specification related in some fashion to the MSC-based requirements must be obtained. The main difficulty here is that the inter-object interactions described in forms of MSCs must be synthesized as executable specifications given in terms of intra-object behaviors as identified as [3]. This is a difficult problem and it has been studied in various limited contexts [3], [4], [5], [6]. A method of using MSCs to construct executable specifications in a more direct way is proposed in [7]. The main point of reference of their work is the formalism of Live Sequence Charts (LSC) [8] in which the component interactions are elaborated in a powerful way using the LSC language while the control flow information is completely suppressed.

In this paper, we mainly contribute to the way of specifications of the protocols (based on MSCs) and to the way of translation of the specifications into a program of formal verification tool for formal verification. The remaining parts of this paper are organized in the following ways. In section 2, we present some related topics such as MSCs, Computation Tree Logic (CTL) that is used by SMV to specify the property of the protocol to be verified and SMV. In section 3, we present the syntax of the specification of the protocols. The translation process is discussed in section 4. In section 5, we present some verification results of the AMBA protocol that we have found through the SMV program generated by the translator. The conclusion and future works are included in section 6.

## 2. Related Topics

In this section, we describe some background knowledge related to our work. We narrate briefly MSCs, CTL and SMV. The definitions and notions discussed in this section will be used in the following sections.

## 2.1 Message Sequence Chart (MSC)

Message Sequence Charts (MSCs) are an attractive visual formalism that is often used in the early stage of system design to specify the system requirements. A main advantage of an MSC is its clear graphical layout which immediately gives an intuitive understanding of the described system behavior [9]. MSCs are particularly suited to describe the distributed telecommunication software [10], [11]. The wide ranges of use of MSCs are usually in the distributed systems and in a number of software methodologies [11], [12], [13]. In a distributed system, MSCs mainly concentrate on the exchange of messages among various processes and their environments as well as some internal actions in these processes. MSCs are also known as object interaction diagrams, timing sequence diagrams and message flow diagrams.

In MSCs, the executing processes are shown by the vertical lines; these processes communicate through an explicit message passing (send-receive) among them shown by the horizontal or downward sloped arrow lines. The head of the arrow indicates the event *message-receiving* and the opposite end indicates the event *message-sending*. Each *send-receive* event (horizontal or downward sloped line) is labeled by the message identifier. For more clear understanding the MSC may also contain necessary data attributes as part of the message exchanged. A simple MSC is shown in the following Fig. 1 where there are two processes namely 'CPU' and 'Memory'.

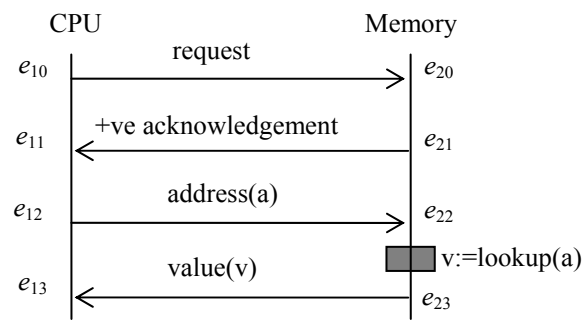

Fig. 1: A Message Sequence Chart

Time flows downward in each vertical line of MSCs. So, in this MSC, the sequences of actions in process 'CPU' and 'Memory' are {sending request, receiving +ve acknowledgement, sending address *a* and receiving value of address *a*} and {receiving request, sending +ve acknowledgement, receiving address *a*, an internal action *v: = lookup(a)* and sending value *v*} respectively. These orderings cannot be violated in either of the processes i.e. a total ordering of the events along every process is assumed. Every process of the MSC is assumed to contain a message queue to store the incoming messages and another message queue to store the outgoing messages. Each MSC is associated with a 'guard'. The MSC is executed when its guard is *true*. The guard is composed by predicates over the variables of the processes in the MSC connected by the logical connectives. The guard of the above MSC is *CPU.status* $\wedge$ *Memory.ready*. The variable *status* is a boolean variable of the process *CPU* which is true when the *CPU* is ready to interact with *Memory*. Similarly, the variable *ready* is a boolean variable of the process *Memory* which is true when it can serve the request of the *CPU.*

Let us discuss MSCs with its formal definition. Assume that $P$ is the finite set of processes, $M$ is the finite set of messages and $A$ is the finite set of internal actions. For each $p \in P$, a set of events the process $p$ takes part in is defined by $\sum_p = \{<p!q, m> \mid p \neq q, q \in P, m \in M\} \cup \{<p?q, m> \mid p \neq q, q \in P, m \in M\} \cup \{<p, a> \mid a \in A\}$. The meanings of $<p!q,m>$, $<p?q,m>$ and $<p,a>$ are 'process $p$ sends message $m$ to process $q$', 'process $p$ receives message $m$ from process $q$' and 'process $p$ performs internal action $a$' respectively. We set $\sum = \cup_{p \in P} \sum_p$ and let $\alpha, \beta$ range over $\sum$. Assume a set of channel $Ch = \{(p,q) \mid p \neq q\}$ and let $c$, $d$ range over $Ch$. A $\sum$-labeled poset is a structure $S = (E, \leq, \lambda)$ where $(E, \leq)$ is a poset and $\lambda: E \rightarrow \sum$ is a labeling function. Here $E$ is a finite set of events and $\leq$ is a partial order which is *reflexive, transitive* and *anti-symmetric*. For any event $e \in E$, $\downarrow e = \{e_1 | e_1 \leq e\}$ where $e_1 \leq e$ means that the event $e_1$ occurs before event $e$. For $p \in P$, and $a \in \sum$, let $E_p = \{e | \lambda(e) \in \sum_p\}$ and $E_a = \{e | \lambda(e) \in a\}$. For channel $c$, let the relation $R_c = \{(e, e1) | \lambda(e) = p!q, \lambda(e_1) = p?q$ and $|\downarrow e \cap E_{p!q}| = |\downarrow e_1 \cap E_{p?q}|\}$. For a process $p \in P$, the relation is $R_p = (E_p \times E_p) \cap \leq$. The $R_{ch}$-edge across the processes is depicted by the horizontal or downward sloped edge.

Thus, an MSC (over $P$) is a finite $\sum$-labeled poset $S = (E, \leq, \lambda)$ that satisfies the following conditions [14]: a) For every $p \in P$, $R_p$ is a linear order. b) For every $\{p, q\} \in P$ and $p \neq q$, $|E_{p!q}| = |E_{q?p}|$ i.e. no lifeless communication edge exists in MSC that means the number of *sent* messages equals the number of *received* messages. c) $\leq = (R_p \cup R_{Ch})^*$ where $R_p = \cup_{p \in P} R_p$ and $R_{Ch} = \cup_{c \in Ch} R_c$ i.e. the partial order of MSC is its visual order; deduced by linear orders of participating processes and the *sent-receive* order of the messages.

The *agents(S)* is the set of agents (processes) taking part in the MSC $S = (E, \leq, \lambda)$ defined as $agents(S) = \{p | E_p \neq \varnothing\}$.

## 2.2 CTL

SMV uses CTL to specify the properties to be verified. In this section, a brief description of CTL is stated.

Atomic propositions, standard boolean connectives of propositional logic and temporal operators all together are used to build the CTL formulae. If $AP$ is a finite set of atomic propositions then- a) $p \in AP$ is a formula, b) if $\varphi$ is a formula then $\sim \varphi$ is also a formula, c) if $\varphi$ and $\varphi_1$ are formulae then $\varphi \vee \varphi_1$ is also a formula, d) if $\varphi$ is a formula then $\mathbf{EX}(\varphi)$, $\mathbf{AX}(\varphi)$, $\mathbf{EF}(\varphi)$, $\mathbf{AF}(\varphi)$, $\mathbf{EG}(\varphi)$ and $\mathbf{AG}(\varphi)$ are also formulae, e) if $\varphi$ and $\varphi_1$ are formulae then $\mathbf{EU}(\varphi, \varphi_1)$ and $\mathbf{AU}(\varphi, \varphi_1)$ are also formulae.

Each temporal operator is composed of two parts: a path quantifier (universal ($\mathbf{A}$) or existential ($\mathbf{E}$)) followed by a temporal modality ($\mathbf{F}$, $\mathbf{G}$, $\mathbf{X}$, $\mathbf{U}$). There are generally many execution paths (the sequences) of state transitions of the system starting at the current state. The path quantifier indicates whether the modality defines a property that should be true for all those possible paths (denoted by universal path quantifier $\mathbf{A}$) or whether the property needs to hold only on one path or on some paths (denoted by existential path quantifier $\mathbf{E}$). The temporal modalities describe the ordering of events in time along an execution path and have the following meanings: a) $\mathbf{F}\varphi$ (read as "$\varphi$ holds sometime in the future") is true in a path if there exists a state in that path where formula $\varphi$ is true, b) $\mathbf{G}\varphi$ (read as "$\varphi$ holds globally") is true in a path if $\varphi$ is true at each and every state in that path, c) $\mathbf{X}\varphi$ (read as "$\varphi$ holds in the next state") is true in a path if $\varphi$ is true in the state reached immediately after the current state in that path, d) $\varphi$ $\mathbf{U}$ $\phi$ (read as "$\varphi$ holds until $\phi$ holds") is true in a path if $\phi$ is true in some state in that path, and $\varphi$ holds in all preceding states.

### 2.2.1 Specification of Properties in CTL

In this section, some examples of common constructs of CTL formula to specify the specifications of the systems in verification are stated. These are the followings: a) $\mathbf{AG}$ ($x{\rightarrow}\mathbf{AF}y$): For all reachable states (AG), if $x$ is asserted in the state, then always at some later point (AF), we must reach a state where $y$ is asserted. b) $\mathbf{AG}(\mathbf{AF}x)$: The proposition $x$ holds infinitely often on every computational path. c) $\mathbf{AG}$ ($x{\rightarrow}\mathbf{A}(x\mathbf{U}y)$): It is always the case that if $x$ occurs in any state, then eventually $y$ is true, and until that time, $x$ must continue to be true. d) $\mathbf{EF}(x{\wedge}\mathbf{EX}x){\rightarrow}\mathbf{EF}(y{\wedge}\mathbf{EX}\ \mathbf{EX}z)$: If it is possible for $x$ to be asserted in three consecutive states, then it is also possible to reach a state where $y$ is asserted and from there after two more steps a state where $z$ is asserted.

### 2.3 Symbolic Model Verifier (SMV)

Symbolic Model Verifier (SMV) [15] is a formal verification tool. It is used for checking finite state systems ranging from completely synchronous to completely asynchronous and from the detailed to the abstracts. In SMV, the Computation Tree Logic (CTL), one kind of temporal logics is used to state the specifications of the system to be verified. The CTL permits a rich class of temporal properties like safety, fairness, liveness etc. to be specified in a concise syntax. SMV verifies the stated specifications investing all the possible behaviors of the system i.e. this is in contrast to a simulator, which only verifies the behavior of the system for the provided vectors.

A SMV specification consists of a collection of properties each of which may be as simple as a statement that a particular pair of signals are never asserted at the same time, or it might state some complex relationship in the values or timing of the signals. SMV allows concise specifications about temporal relationships between signals, and can automatically be verified. SMV uses the Binary Decision Diagram (BDD)-based symbolic model checking algorithm to effectively and efficiently find out if the system specifications are satisfied or not. If a specification is not satisfied by the model, SMV automatically produces a counterexample. For this, SMV is a very effective debugging tool as well as a formal verification system.

### 3. Specifications of the Protocol

The specifications is composed of a finite set of processes each of which is a system component and performs a list of *transaction schemes* where a *transaction scheme* is the unit of interactions among different processes and consists of a guarded choice between a set of *transactions* where a *transaction* is modeled as an MSC.

The protocol is described by a textual input file and the target is to build the corresponding SMV program for it to be used for verification. The syntax for specifying the protocol and mapping of these specifications to the target SMV file are narrated below. Fig. 2 shows the syntax.

| | |
|---|---|
| S | → **PROTOCOL protocol-name {** proc-dec-part   trans-sch-dec-part **}** |
| proc-dec-part | → proc-dec-part  proc-dec \| proc-dec |
| proc-dec | → **PROCESS process-name {**var-dec-part    equation-part   **}** |
| type | → basic-type \| enum-type \| array-type |
| basic-type | → range-type \| **BOOLEAN** |
| enum-type | → **{** id-list **}** |
| id-list | → **identifier (, identifier)**$^+$ |
| array-type | → **ARRAY** range-type **OF** basic-type |
| range-type | → **interger-const .. interger-const** |
| var-dec-part | → var-dec-part var-dec**;** \| ε |
| var-dec | → **type-id :** id-list |
| equation-part | → **EQUATION** id-list**;** |
| trans-sch-dec-part | → trans-sch-dec trans-sch-dec-part \| trans-sch-dec |
| trans-sch-dec | → **SCHEME trans-schm-name {** trans-list **}** |
| trans-list | → trans-dec trans-list \| trans-dec |
| trans-dec | → **TRANSACTION trans-name {** <br>    **AGENTS {** agent-list **}** guard-section **; }** <br>    **\| TRANSACTION trans-name { AGENTS {** agent-list **}** |
| guard-section | → **GUARD** guard **;** |
| agent-list | → agent-list agent \| agent |
| agent | → **process-name :** event-list |
| event-list | → event **,** event-list \| event **;** |
| event | → **send(mesg-id, var) \| send(mesg-id, const, type) \|** <br>    **recv(process-name.mesg-id)** \| {action} |
| action-atom | → simple-stmt **;** \| if-stmt |
| action | → action action-atom \| action-atom |
| simple-stmt | → **var :=** expr \| **var :=DIN** |
| expr | → expr [**\*** \| **/** \| **&**] F \| F |
| F | → F **+** G \| F **-** G \| F \| G |
| G | → G **mod** H \| H |
| H | → H **RelOp** I \| I |
| I | → ~I \| -I \| **(**expr**)** \| var \| const |
| const | → **integer-const \| boolean-const** |
| guard | → guard **&** guard-atom \| guard-atom \| (guard-atom) |
| guard-atom | → ~prop \| prop |
| prop | → prop **or** prop-atom \| prop-atom |
| prop-atom | → scoped-var relop const \| scoped-var relop scoped-var \| scoped-var |
| relop | → **=** \| **<** \| **>** \| **≤** \| **≥** \| **!=** |
| if-stmt | → **IF** expr **{** action **}** \| **IF** expr action-atom \| **IF** expr **{**action**}ELSE {**action**}** |
| var | → **identifier \| identifier[identifier] \| identifier [integer-const]** |
| scoped-var | → **process-name.**var |
| **protocol-name** | → **identifier** |
| **process-name** | → **identifier** |
| **trans-name** | → **identifier** |
| **mesg-id** | → **identifier** |

Fig.2: Syntax of the specifications

At this point, let us describe the input file with the following simple example shown in Fig. 3.

```
protocol cpu_bus_mem {
        process cpu {    //process declaration
                0..7 : addr_buf, data_buf;    //range type variable
                boolean: status;
                equation T1;    //transaction scheme
        }
        process bus {    //process declaration
                ……………….. //variables and transaction scheme declaration
        }
        process mem{    //process declaration
                …………….. //similar
        }
        scheme T1 {    //start of transaction scheme T1
                transaction transfer {    //one transaction 'transfer'
                        agents {    // agents of this transaction ('cpu' and 'bus')
                                cpu : send(req,1,boolean),    // send event
                                        recv(bus.ack),    //recv event
                                        {status:=din;};    //internal action
                                bus : recv(cpu.req),
                                        send(ack,0,boolean);
                        }
                        guard  ~bus.ready  & cpu.status;    //guard for this transaction
                }    //end of this transaction
                transaction not_transfer {
                        …………………. //similar
                }
        }    //end of transaction scheme T1
        scheme T2{    //start of transaction scheme T2
                …………...
        }    // end of transaction scheme T2
}    //end of input file
```

Fig. 3: An Example of the Specifications

The file starts with the keyword *protocol* followed by the name of the protocol (cpu_bus_mem in the example). The processes are the components of the protocol considered. The 'cpu', 'bus' and 'mem' are the processes in the 'cpu-bus-mem' example. The variables declaration part of each process contains the variables which are used to store the messages received from other processes and also the local variables to perform internal actions. The types supported are *boolean*, *range*, *enumeration* and arrays of these basic types.

The 'equation' of a process describes the order in which it will take part in various transaction schemes and is declared in the input language using the *equation* construct. Thus we assume that in this restricted version of our work, the control flow within each process is *cyclic*.

The Transaction Scheme part contains the description of each transaction scheme and is described using the *scheme* construct. Each scheme consists of one or more transactions where each of the transactions is modeled as a guarded MSC. The transaction executes only if its guard is true.

In each transaction the participating processes are declared using the *agents* construct. The actions of each process in this transaction are also defined here. These actions can be sending or receiving of messages or internal actions as described below.

- A *send* action consists of a message id (which should be unique both in sender and receiver), the value to be sent (can either be a boolean/integer constant or a variable of suitable type) and the data-type of value. The data-type is required in the *send* declaration in order to determine which port the data is being sent on. As for example, the integer value 1 can represent a boolean value or a value of type *range[n1..n2]*. This allows us to enforce type matching on the sender and receiver ends i.e. if a process $p$ sends a message $x$ and a process $q$ receives this message as $y$, it is expected that the data-types of $x$ and $y$ to be the same.

- A receive declaration *recv* specifies message-id and the process from which the message is received. One special variable called *din* stores the value of the messages received. This is necessary for the local resolution of the conflicts in the processes. This variable is the key to get all the necessary values from other processes to the current process to decide on the resolution condition. The assumption here is that a process receives all the variable values through *din* to decide on a resolution condition. The variable *din* is *struct type* which has a field for each of the data-types used in the specification file. It also contains a field for storing the data-type of the last message received. A process can use the value of a received message by assigning *din* to a variable. Since there is only one *din* variable per process, it should be made sure to use the received value before it is over-written by another 'receive' message.

- An internal action is specified by enclosing a set of sequential statements (like checking of a logical condition, assigning a value to a variable etc.) within parenthesis. Needless to say, an internal action should involve only the local variables of the process and the special variable *din*.

The guard of each transaction consists of propositions of all the processes in that transaction. In a guard, it is not expected to compare the values of variables of different processes (variables within same process can be compared). It is expected that all the transactions within a transaction scheme can be distinguished for a particular process by the variable/variables of that process in the guards of the transactions.

Now we want to show how we can represent the transaction within a transaction scheme of a bus protocol using MSCs. We have taken AMBA bus protocol as an example. In AMBA protocol, these are master component $P_m$, interface of the master $I_m$, the bus controller $BC$, interface of the slave component $I_s$ and the slave component $P_s$. The master $P_m$ sends data to the slave $P_s$. The master $P_m$ enqueues data into the queue of the master interface $I_m$ which requests the bus controller $BC$ for the bus access. Getting the access, if the slave interface $I_s$ is ready, $I_m$ sends the data to the slave interface which finally dequeues data to the slave $P_s$. The following Fig. 4 shows four transaction schemes where the scheme in figure (a), (b), (c) and (d) show the *enqueuing* of data, requesting for the bus access and the normal data transfer and *dequeuing* scenarios. The formula stated at the bottom of each MSC is the guard for that MSC.

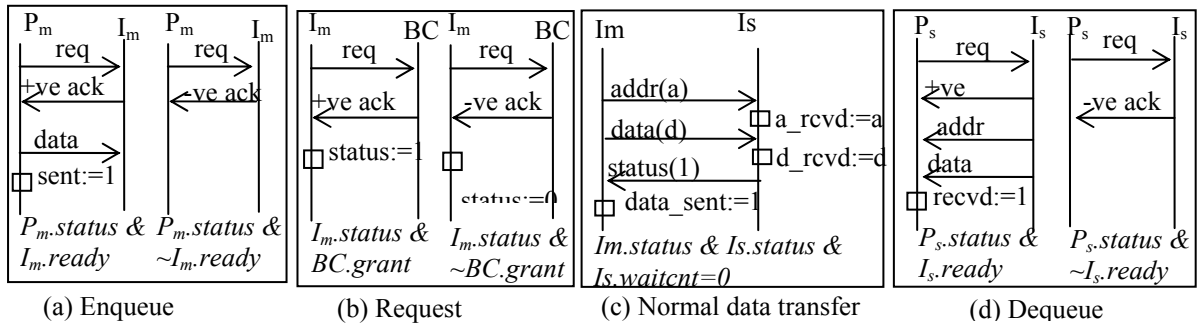

Fig. 4: Specification of some AMBA features using MSCs

The transaction scheme *Enqueue* (Fig. 4(a)) shows the enqueuing data from the master $P_m$ into the queue of the master interface $I_m$. This scheme happens when the *status* of the master $P_m$ is true. When the master interface component is ready to accept data from the master then a successful enqueuing i.e. the left transaction (MSC) occurs and an internal action shown by the small box on the life line of $P_m$ happens; otherwise the right transaction (enqueuing unsuccessful) occurs. The scheme *Request* (Fig. 4(b)) involves $I_m$ and *BC*. A request made for the access of the bus by the master interface to the bus controller *BC*. When the variable *status* of the master interface $I_m$ is true then $I_m$ requests the bus for the bus access. If *BC.grant* is true i.e. if the bus controller can grant the access to the master interface then $I_m$ sets true to *status* (left MSC) else false is set to *status* (right MSC). Let us now describe scheme *Transfer* (Fig. 4(c)). In this scheme, the participating processes are $I_m$, $I_s$ and *BC*. Before describing the transactions in this scheme, let us present some local variables of these processes. The variable *data_sent* holds when the data is sent. The *maxwait* is the variable that is fixed and indicates after how many wait cycles the bus access to the master will be suspended. The variable

*waitcnt* holds the number of wait cycles encountered. The variables *a_rcvd* and *d_rcvd* hold the address and data respectively received from $I_m$. When the variable *status* of $I_m$ holds (it has bus access) and the variable *status* of $I_s$ holds ($I_s$.*status* holds when $I_s$ can accept data/address from $I_m$ i.e. its queue is not full) and *waitcnt* is 0, then the normal data transmission occurs. Here the life line of *BC* is not shown as there is only internal action of it and that is simply a no-op. In contrast to the scheme *Enqueue* the scheme *Dequeue* (Fig. 4(d)) denotes the dequeuing of data from the queue of the slave interface $I_s$ by the slave $P_s$. When the variable *status* of the slave is *true* then it can get or cannot get the data and address from the slave interface depending on the value of the variable *ready* in $I_s$.

Now let us see how we can map the specifications with the target SMV program. One module for each of the processes in the protocol is created in the SMV file. The variables declared for a process in the input file are also in the corresponding module of that process in the SMV file. It is assumed that there are two queues for every ordered pair of processes which interact between themselves ($p$ sends messages to $q$ and also receives messages from $q$, so there are two queues for them). The sender enqueues messages and the receiver dequeues messages. A queue size ($Q\_SIZE$) is defined and fixed for all processes. For the module of the process *p*, some boolean variables such as *full_p_q*, *empty_q_p* and some *range type* variable namely *tail_p_q, head_q_p* are declared where $p \neq q$ and $q \in \{$set of all processes in the protocol$\}$. The range of these *range type* variables is from 0 to $Q\_SIZE$-1. The initial values of the variables *head_q_p* and *tail_p_q* are set to zero. Every time when a message is received by process *p* from *q*, the next value of the variable *head_q_p* is set to (*head_q_p*+1) *mod* $Q\_$SIZE and when a message is sent by *p* to *q* then the next value of variable *tail_p_q* is set to (*tail_p_q*+1) *mod* $Q\_$SIZE. The variable *full_p_q* holds (*((tail_p_q+1) mod* $Q\_SIZE = head\_p\_q$)) when the queue (for sending message) between *p* to *q* is full and hence no message can be sent further from *p* to *q* before the process *q* receives the message from *p*. Similarly, when *empty_q_p* holds ( *head_q_p = tail_q_p* ) then the process *p* has nothing from *q* to receive. A scalar type variable *state* (*state*: = {_p0, _p1, _p2, ………, _pn}) is declared for each process *p* which contains all the states that the process *p* can be in. The value of *n* is calculated from the total number of different events the process *p* takes part in.

# 4. Translator

In the previous section, we discuss the syntax of the specifications with example and the mapping of the specification to the target SMV file. Now it is the time to look at the translation process of the specifications into SMV program. The first step is the standard lexical analysis. This is followed by a parsing step checking syntax. After the confirmation from the parser that the input file is grammatically right, the translation to the target code (SMV) starts. We briefly describe the lexical analyzer, syntactic analyzer (parser) and generation of SMV code in this section.

## 4.1 Lexical Analysis

JFlex is the lexical analysis tool used in the translator. In addition to the lexical analysis, JFlex called by the parser (in this case the parser is Constructor of Useful Parser (CUP) which is paired with JFlex) does tasks such as i) the identification code with the token type of the recognized token is passed to parser, ii) the value of the integer literal and other information is passed to parser and iii) does the programmer coded actions after recognition of the token. The first of these three tasks is used for syntactic analysis discussed in the next section and the other two tasks help the parser to generate the target code.

The valid lexical tokens of the specifications are written in the file 'lexer.flex' from which a java version ('MyScanner.java') of the lexical analyzer (also known as scanner) is created by the command *'jflex lexer.flex'*. This 'MyScanner.java' is the scanner or the lexical analyzer for the specification file.

The 'lexer.flex' file is composed of three parts such as i) user code, ii) options and declarations and iii) lexical rules each divided by the sign *%%*. Let us see the first section 'user code'. In this section, the *package* and *import* statements are written that should be exactly at the top of the generated scanner class ('MyScanner') i.e. the text up to the first line starting with *%%* (starting of the second section) is copied verbatim to the top of the scanner class. In our case, this section contains i) *import java_cup.runtime.*;* ii) *import JFlex.*;* and iii) *import java.util.*;* The second section 'options and declarations' is more interesting. It consists of a set of options, code that is included inside the generated scanner class ('MyScanner'), macro declarations and lexical states. The last section 'lexical rules' contains the regular expressions and actions that are to be executed in the case when the scanner matches the associated regular expression.

## 4.2 Syntactic Analysis

The syntactic analyzer checks if the file is organized according to the grammar once the lexical analyzer assures that all the tokens are valid in the source file. For example, the lexical analyzer may pass the five valid tokens such as *guard { transaction agents }* to the parser. But the parser sees it as invalid organization.

As stated earlier, CUP is used as the parser for the translation. The CUP with its usual task, does one additional job and that is it can perform any code the programmer wants to encode upon recognizing a valid grammatical construct. This helps generating target code in two ways i) the generated code is written to a file to be executed later and ii) the generated code is executed during parsing.

The syntactic structure of the specification file is written in the file 'grammar.cup' from which some files are produced by the command *java java_cup.Main < grammar.cup*. On of these files is *parser.java* and it is the syntactic analyzer (parser) for our specification file. There are five sections in a specification such as i) package and import specifications, ii) user code, iii) terminal and non-terminal lists, iv) precedence and associativity of terminals and v) grammar. Each of these parts must appear in order. We have not discussed these sections here.

## 4.3 Generating SMV Code

After the scanning and parsing, the final task for the translation process is to create the target code i.e. SMV code. The language used for this task is Java.

Let us look at the *Main* class of the implementation. Before starting the class *Main*, the class *java.io.\** is imported. The *Main* class is then declared. In this class, the method *main* is declared where the parser is called to analyze the specifications (input file) syntactically. The parser then calls the scanner that analyzes the input lexically, at the time the parser requires the next lexical token of the input file. This is done in the *try* block shown below:

try{ parser p = new parser(new MyScanner(new FileReader(argv[0]))); p.parse();}

The *try* block means that if something fails then the program exits that block. The first line of the *try* block creates a new parser object and the second line starts the parser.

In the *main* method there is a *catch* block after the *try* block. This *catch* block takes the exception, the reason why *try* block fails and to clean up errors occurred before the program quits. This is the completion of the *Main* class.

Several classes are designed and implemented to generate the SMV code. Among them the following are some important classes.

i) Protocol.java: An object of this class represents the protocol.

ii) Scheme.java: An object of this class represents a transaction scheme of the specifications.

iii) Process.java: An object of this class represents a process of the protocol. It holds all the local variables and *equation* of this process and the methods that are called by the Protocol.java to generate the SMV code for this process.

iv) Transaction.java: An object of this class represents a transaction in a transaction scheme.

v) Node.java: An object of this class represents the node in the tree where

a tree is a different representation of the events of a process that gives the sequence in which the SMV code for a process has to be generated.

Now let us look at the way of creating the SMV code for a process briefly. A separate event sequence is generated for each process for each scheme, and these event-sequence are then concatenated together according to the equation of each process. Suppose the event sequence for the process $P$ for the scheme $S$ is to be generated. Let the transactions of $S$ in which $P$ is an agent be $Tp1, Tp2, \ldots Tpn$. For generating the event sequence, each transaction is projected (accumulating all its actions and guards) onto $P$. Thus a list of propositions $Gp1, Gp2, \ldots Gpn$ (the guard of this process is simply the conjunction of these propositions) and a list of event $Ep1, Ep2, \ldots Epn$ are achieved.

Now the leading (first event in the list) event of each transaction (projected on $P$) is examined and the transactions with same event are grouped into one list. If there is more than one group i.e. there exist at least two transactions which differ in the leading event then it is understood that the transactions are not isomorphic. Hence these groups are labeled as 'positive' and 'negative' based on the leading atomic proposition of their guards. For instance, suppose the groups $\{Tp1, Tp3\}$, $\{Tp4\}$ and $\{Tp2, Tp5\}$ are made by comparing the leading event. Suppose the leading atomic proposition is $p$ (or $\sim p$). Then each transaction in a particular group should have the exactly same sign of the proposition. Furthermore, different groups with the same sign of $p$ are merged together and the whole process is repeated i.e. if the leading proposition were $\{p.v1, \sim p.v1, p.v1, p.v1, \sim p.v1\}$ in the above example, then the new groups will be $\{Tp1, Tp3, Tp4\}$ and $\{Tp2, Tp5\}$. In the generated code, value of variable $v1$ will be checked at this place to determine whether the next event should be as dictated by $\{Tp1, Tp3, Tp4\}$ or $\{Tp2, Tp5\}$.

If all the transactions initially have the same leading event, then their second events are examined in order to form groups and so on. Thus the algorithm consists of two distinct steps.

- Identify the point where event sequences of transactions begin to differ.
- Identify the proposition based on which the process will make a local decision as to which transaction is to be followed. We may need to look at more than one proposition to uniquely identify the transaction.

The states in translated code correspond to each action of a process. As an optimization, a sequence of actions for which no guards have to be evaluated in between, can be collapsed into one state.

In the translated SMV code, message passing is handled using bounded length queues (whose length can be given as a parameter during translation phase). The blocking semantics is followed in case of full or empty queues.

## 5. Verification Result

In this section, we present some verification results of the AMBA bus protocol (described in section 3) using SMV program generated by the translator. Each specification is followed by the meaning of the specification which is followed by the SMV result.

Specification 1: *EF BC.grant*
Meaning: This specification states that there exists a computation path where in some future state *BC.grant* holds. This specification checks if the bus controller can grant the bus access to the master in some future time in at least one execution.

SMV Result: This specification is *true*.

Specification 2: *EF I$_s$.status*
Meaning: This specification states that there exists a path where in some future state *I$_s$.status* holds. This specification checks if the slave interface *I$_s$* is capable of receiving data and address from the master interface *I$_m$* in some future time.
SMV Result: SMV shows that this specification is *true*.

Specification 3: *AG (P$_m$.sent → AF (P$_s$.recvd))*
Meaning: This specification states that in every reachable state if the master *P$_m$* sends data then eventually the data will be received by the slave *P$_s$*.
SMV Result: SMV shows that this specification is *true*.

## 6. Conclusion and Future Works

We have presented our work on a high level executable specification mechanism for specification of bus protocols that combines distributed control flow features with interactions refined as Message Sequence Charts (MSCs). In this chapter, we highlight the summary of our work and point some future research directions.

### 6.1 Summary of Our Work

We have contributed to the following main tasks:

- The way of specification of the protocol using MSCs.
- Construction of a translator that translates the specifications into SMV programs. The SMV is the well known model checking tool based on CTL. These SMV programs then formally verify the properties of the protocol.

### 6.2 Future Work

We see many further possibilities to continue this work. Some of the important works that may be investigated in future are the following:

- The current restriction on the guards of the transactions within a transaction scheme can be removed for more flexible way of specifying the specifications.
- The states for a process in translated code of SMV program correspond to each action of that process. As an optimization, a sequence of actions for which no guards have to be evaluated in between, can be made into one state.
- Another interesting issue to pursue is to translate the protocol into another powerful model checking tool SPIN. The specifications can also be translated into Hardware Description Language (HDL) like verilog or VHDL.